\documentclass[prb,twocolumn,tbtags,superscriptaddress,floatfix]{revtex4-2}

\usepackage{amsmath}
\usepackage{amssymb}
\usepackage{graphics}
\usepackage{epsfig}
\usepackage{verbatim}
\usepackage{textcomp}
\usepackage[colorlinks=true,urlcolor=blue,linkcolor=blue,citecolor=blue]{hyperref}
\usepackage{tabularx}
\usepackage{xcolor}
\usepackage{braket} 
\usepackage{float}
\usepackage{comment}

\begin{document}
\title{Efficient implementation of quantum signal processing via the adiabatic-impulse model}

\author{D.~O.~Shendryk}
\thanks{These authors contributed equally}
\affiliation{B.~Verkin Institute for Low Temperature Physics and Engineering  of the National Academy of Sciences of Ukraine, Kharkiv 61103, Ukraine}
\affiliation{Ruhr-Universität Bochum 44780, Germany}
\author{O.~V.~Ivakhnenko}
\thanks{These authors contributed equally}
\affiliation{B.~Verkin Institute for Low Temperature Physics and Engineering  of the National Academy of Sciences of Ukraine, Kharkiv 61103, Ukraine}
\affiliation{RIKEN Center for Quantum Computing, RIKEN, Wakoshi, Saitama, 351-0198, Japan}
\author{S.~N.~Shevchenko}\email{sshevchenko@ilt.kharkov.ua}
\affiliation{B.~Verkin Institute for Low Temperature Physics and Engineering  of the National Academy of Sciences of Ukraine, Kharkiv 61103, Ukraine}
\author{Franco~Nori}
\affiliation{RIKEN Center for Quantum Computing, RIKEN, Wakoshi, Saitama, 351-0198, Japan}
\affiliation{Quantum Research Institute and Physics Department, The University of Michigan, Ann Arbor, MI 48109-1040, USA}
\begin{abstract}
 We  investigate  an  analogy between quantum signal processing (QSP) and the adiabatic-impulse model (AIM) in order to implement the QSP algorithm with fast quantum logic gates. QSP is an algorithmic technique that uses single-qubit dynamics to perform 
a polynomial function transformation. 
 The adiabatic-impulse model  effectively describes the evolution of a two-level quantum system under  a  strong external driving field. 
We can map parameters from QSP to  the  AIM to implement  a  QSP-like evolution with non-adiabatic, high-amplitude external drives. By choosing  the  AIM  driving  parameters that control  the  non-adiabatic transition parameters (such as driving amplitude~$A$, frequency $\omega$, and signal timing), one can achieve polynomial approximations and increase robustness in quantum circuits. 
 The analogy between QSP and AIM  presented here  can be useful as a way to directly implement the QSP algorithmic technique on quantum systems.  And show the  benefits from fast Landau-Zener-Stückelberg-Majorana (LZSM) quantum logic gates  in comparison with usual resonance driving gates in IBM-Q quantum processors.
\end{abstract}

\maketitle
\newpage

\section{Introduction}

Quantum algorithms represent an innovative paradigm in computational science, using the principles of superposition, entanglement, and quantum interference to confront overwhelmingly time-consuming problems for classical approaches \cite{shor1994, QuantumSearchGrov,QuantumWalk, QuantumSearch,HamiltonianSimulation,Nielsen2012,Low2019hamiltonian,Li2022}. 
 Quantum algorithms have been reinterpreted within more general frameworks that emphasize common structural principles. A notable example is the perspective framework,  developed in Ref.~\cite{martyn2021grand}, named  \textquotedblleft grand unification of quantum algorithms\textquotedblright, which shows that a wide range of fundamental algorithms can be understood as instances of singular value transformations. At the core of this view lies  Quantum Signal Processing (QSP), a method exploiting single-qubit dynamics with interference, enabling polynomial function transformations with $O(d)$ elementary unitary operations on the quantum subsystem, where $d$ is the degree of the polynomial function.  The role of QSP as a unifying primitive has motivated its integration into increasingly sophisticated algorithmic designs \cite{GQSP, HamiltonianSimulation, rossi2023quantum}, where it functions both as a direct tool and as a foundation for broader constructions such as Quantum Singular Value Transformation (QSVT). 

QSP involves a combination of adjustable rotations of  a qubit state  around the $z$-axis, which produces  phase differences between energy levels, and fixed $x$-axis rotations on the Bloch sphere, which change the energy levels occupation. These $x$- and $z$-axes rotations are applied one by one in a sequence. Such QSP sequences of pulses can be used  in  a composite-pulse approach that allows to increase robustness to  fluctuations in the driving signal~\cite{Bluvstein2022}. 

We found that the combination of consecutive rotations around the $x$- and $z$-axes is also directly  used in  the so-called adiabatic-impulse model (AIM) \cite{Damski2006}, a model specifically developed for non-adiabatic quantum dynamics  of qubits, with particular emphasis on interferometry \cite{SHEVCHENKO_2010,Ivakhnenko2018,Bonifacio2020, Ivakhnenko_2023}. It is a useful tool to investigate analytically the dynamics of quantum systems with a small energy gap  \cite{Ivakhnenko_2023,Ryzhov_2024,Wen2020}. It combines adiabatic evolution, where the energy levels occupation is conserved but a phase difference is gained, and diabatic (impulse-type) transitions, during which the interference of the amplitudes of different states occurs. 

We find it beneficial to explore the analogy between the two approaches—QSP and AIM—and demonstrate that  the  AIM can be used to effectively implement QSP.  This is possible by utilizing alternative fast non-adiabatic gates \cite{Campbell2020}. 

The rest of the paper is organized as follows. In Sec.~\ref{QSP_and_AIM_Overview}, we review the key notions of  this  work: QSP and AIM. In Sec.~\ref{QSP_AIM_mapping}, we describe the algorithms for converting parameters from QSP to AIM. In Sec.~\ref{Double_LZSM}, we show the way to use two  consecutive  Landau-Zener-Stückelberg-Majorana (LZSM) transitions \cite{SHEVCHENKO_2010,Ivakhnenko_2023,Glasbrenner2023} to implement QSP  faster. In addition, we show an analogy between double LZSM transition and a Mach-Zehnder interferometer to implement a rotation operator around the $x$-axis by an arbitrary angle.

\section{REVIEW OF quantum signal processing AND adiabatic-impulse model}
\label{QSP_and_AIM_Overview}
\subsection{Quantum Signal Processing}

Quantum Signal Processing is a quantum algorithmic technique initially designed for implementing polynomial transformations of matrix eigenvalues via phase modulation, originally  inspired by composite-pulse techniques from nuclear magnetic resonance \cite{WIMPERIS1994221}.  The possibility to implement such polynomial transformations is particularly important in the context of applications like Hamiltonian simulation, as we will demonstrate later.  It can also be used to enhance signal sensitivity to specific input parameters, making it a powerful tool for error mitigation, quantum optimization, and quantum machine learning tasks~\cite{kikuchi2023realization,liao2024machine}. For instance, the technique can be applied to amplify weak signals beyond the limits of repetition statistics and suppress noise without direct measurement \cite{WIMPERIS1994221}.

The QSP algorithm involves using a sequence of unitary rotations of a qubit state on the Bloch sphere around the $x$- and $z$-axes.
The operator responsible for rotations around the $x$-axis is denoted as the \textit{signal operator} $W$, and the operator responsible for rotation around the $z$-axis is called the \textit{signal processing operator} $S$
\begin{eqnarray}
	&W(a)& = \exp{\left(i\frac{\theta(a)}{2} X\right)} =
	\begin{pmatrix}
		a & i\sqrt{1 - a^2} \\
		i\sqrt{1 - a^2} & a
	\end{pmatrix},\notag\\
	&S(\phi)&  = \exp{(i\phi Z)},
\end{eqnarray}
where $X$ and $Z$ stand for the Pauli matrices $\sigma_x$ and $\sigma_z$, respectively. Here, the parameter $a$ is defined as
\begin{align}
	a = \cos(\theta/2) \  \in \  [-1,1].
	\label{Polinomialbase}
\end{align}
For a set of phases
\begin{equation}
	\boldsymbol{\Phi} = (\phi_0, \phi_1, \dots, \phi_d),
\end{equation} the operator $U$, which represents the evolution in the QSP technique, is defined as the sequential application of the signal processing operator $S$ and the signal operator $W$:
\begin{align} \label{QSP_evolv}
	&U_{\boldsymbol{\Phi}, a} = \exp{(i\phi_0 Z)} \prod_{k=1}^{d} W(a) \exp{(i\phi_k Z)} \in SU(2). 
\end{align}


The operators $W$ and $S$ can be written in the following way using rotation operators
\begin{align}
	W(a(\theta)) = R_x(-\theta), \hspace{2em} S(\phi) = R_z(-2\phi).
\end{align}
 The  quantum rotation gates $R_x$ and $R_z$  describe rotations of single-qubit states on the Bloch sphere by angles $\alpha$ and $\beta$ around the $x$- and $z$-axes, respectively. These gates are defined both in terms of matrix exponentials and explicit matrix representations
\begin{subequations} 
	\begin{align}
		R_x(\alpha) = \exp{\left(-i\frac{\alpha}{2} X\right)} =
		\begin{pmatrix}
			\cos\left(\frac{\alpha}{2}\right) & -i\sin\left(\frac{\alpha}{2}\right) \\
			-i\sin\left(\frac{\alpha}{2}\right) & \cos\left(\frac{\alpha}{2}\right)
		\end{pmatrix}, \\
		R_z(\beta) = \exp{\left(-i\frac{\beta}{2} Z\right)} =
		\begin{pmatrix}
			\exp{(-i\frac{\beta}{2})} & 0 \\
			0 & \exp{(i\frac{\beta}{2})}
		\end{pmatrix}.
\end{align}\end{subequations} 
 Consequently,  the equation~(\ref{QSP_evolv}) can be written in terms of rotation operators $R_x$ and $R_z$
\begin{align}\label{QSP_rot}
	U_{\boldsymbol{\Phi}, a} = R_z(-2\phi_0) \prod_{k=1}^{d} R_x(-\theta) R_z(-2\phi_k).
\end{align}
One of the most remarkable properties of the QSP framework is its ability to construct quantum circuits that implement a desired polynomial transformation on the eigenvalues of an operator. 

The inverse of this principle is also true: given a target polynomial $M(a)$, there  always  exists a corresponding set of QSP phase angles $ \boldsymbol{\Phi}$ such that the transformation $M(a)$ can be exactly represented as $\langle 0 | U_{\boldsymbol{\Phi},a} | 0 \rangle$.
This principle is formalized in Theorem~1 (Quantum Signal Processing) of Ref.~\cite{martyn2021grand}, which states that the QSP sequence $U_{\boldsymbol{\Phi}, a}$ generates a matrix elements, expressible via  a  polynomial function of $a$:
\begin{align}
	U_{\boldsymbol{\Phi}, a} =
	\begin{pmatrix}
		M(a) & iQ(a) \sqrt{1 - a^2}\\
		iQ^*(a) \sqrt{1 - a^2} & M^*(a)
	\end{pmatrix}.
\end{align}
For this construction to hold,  the desired $ M(a)$ and complementary $ Q(a)$ polynomials  must satisfy several conditions. The degree of the polynomial $ M(a) $ must be at most $ d $ and the degree of  the polynomial  $ Q(a) $ must be at most $ (d-1)$.  These  polynomials must also exhibit parity relationships with the degree $ d $, and a normalization condition must ensure the unitarity of the resulting transformation.
\begin{figure}[t] 
	\centering
	\includegraphics[width=0.5\textwidth]{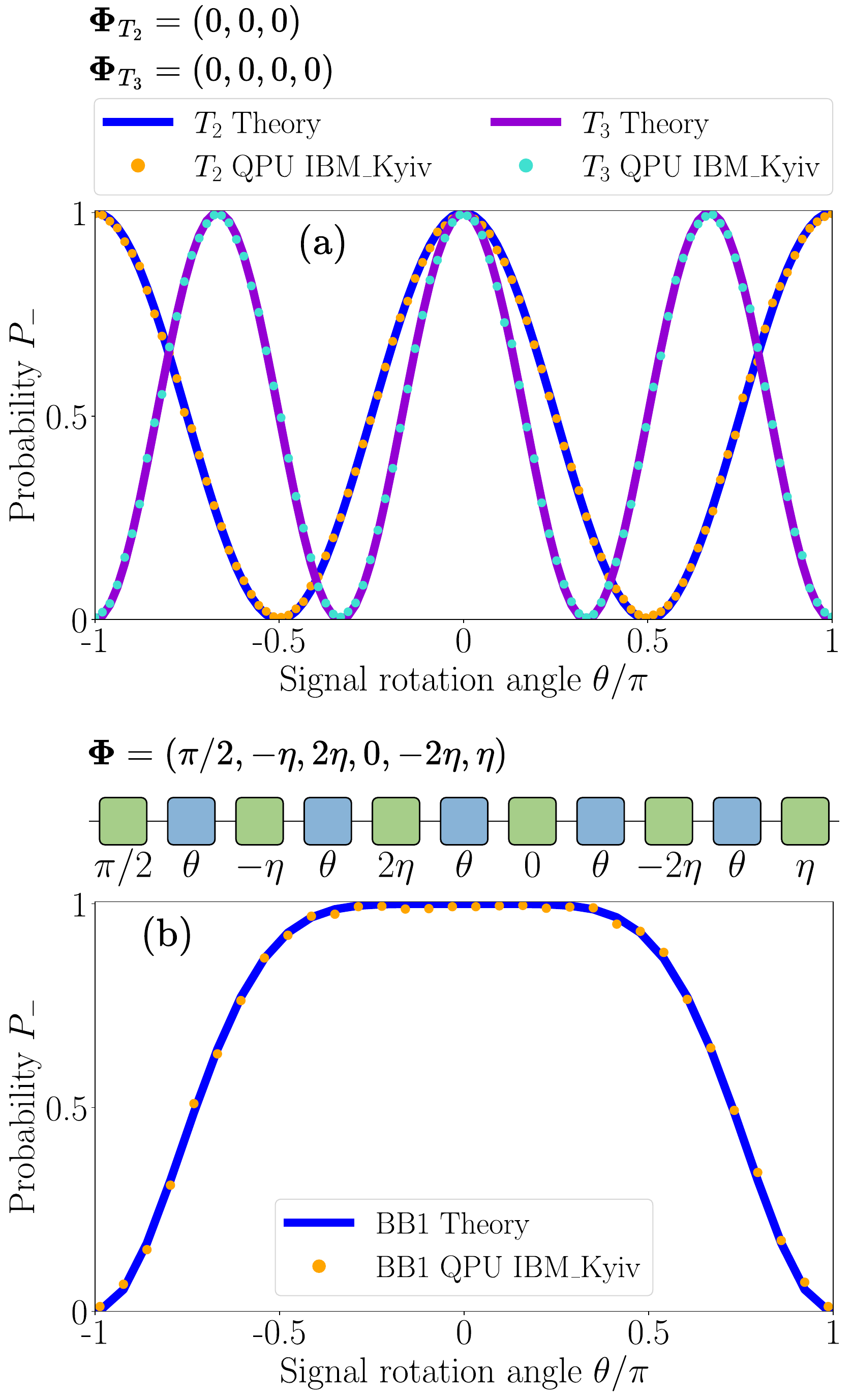}
	\caption{Occupation probability $P_-$ after the QSP sequence  is  applied to the qubit with ground state initial conditions. Chebyshev polynomials with Eqs.~(\ref{Cheb2},\ref{Cheb3}) are shown in panel~(a).  The BB1 sequence, corresponding to the polynomial in Eq.~\eqref{BB1Plynom}, is shown in panel (b) (solid curve).  The dots  in (a,b)  correspond to the probability obtained by performing the respective QSP sequences on an IBM-Q quantum processing unit (QPU). The scheme for the BB1 sequence is shown  (b) in  terms of QSP operators, where  the  green boxes correspond to the signal processing operator $S(\phi)$ and blue boxes correspond to the signal operator $W(a)$.}\label{BB1_fig}
\end{figure}

As an example, we now consider  the  Chebyshev polynomials $T_d(a)$ \cite{abramowitz1965handbook}, which are defined as follows
\begin{subequations} 
	\begin{align}
		& T_0(a) = 1, \hspace{1em} T_1(a) = a; \\
		& T_d(a) = 2aT_{d-1}(a)-T_{d-2}(a) = \cos(d \arccos a),
\end{align}\end{subequations} 
 and how they  naturally arise within the framework of QSP \cite{martyn2021grand}.  Such polynomials play a central role in Hamiltonian simulation, since time-evolution operators of the form $\exp{(-iHt)}$ are often approximated through Chebyshev polynomial series. For trivial phase sequences (i.e., all phases in $\boldsymbol{\Phi}$  equal  zero), the resulting polynomials $ M(a) $ align exactly with $ T_d(a) $:
\begin{subequations} 
	\begin{itemize}
		\item For $ \boldsymbol{\Phi} = (0, 0) $, the corresponding polynomial is 
		\begin{equation} \label{Cheb1}
			M(a) = a = T_1(a);
		\end{equation}
		\item For $ \boldsymbol{\Phi} = (0, 0, 0) $, the polynomial becomes 
		\begin{equation} \label{Cheb2} 
			M(a) = 2a^2 - 1 = T_2(a);
		\end{equation}
		\item For $ \boldsymbol{\Phi} = (0, 0, 0, 0) $, the result is 
		\begin{equation}\label{Cheb3} 
			M(a) = 4a^3 - 3a = T_3(a).
		\end{equation}
\end{itemize}\end{subequations}

Another important QSP sequence is BB1 (Broadband~1) \cite{WIMPERIS1994221}. It is a robust composite pulse sequence used in NMR to correct amplitude errors in RF pulses. This  is particularly useful for achieving high-fidelity spin manipulations, even when the RF field strength is not perfectly calibrated.  Following Ref.~\cite{martyn2021grand} we consider BB1 with a phase sequence 
\begin{eqnarray} \label{BB1_seq}
	\boldsymbol{\Phi} = (\pi/2, -\eta, 2\eta, 0, -2\eta, \eta),\\
	\text{where } \eta = \frac{1}{2} \cos^{-1}\left(-\frac{1}{4}\right). \notag
\end{eqnarray}
 Via the QSP technique, Eq.~(\ref{QSP_evolv}), one can obtain the polynomial form from the phase sequence in Eq.~(\ref{BB1_seq}) 
\begin{equation}
	M(a)=\frac{1}{8}a^2\left[3a^8-15a^6+35a^4-45a^2+30\right]. \label{BB1Plynom}
\end{equation}
The BB1 sequence  in Eq.~\eqref{BB1_seq}  describes a qubit that remains in the excited state in the broad region $\theta\approx (-2/3\pi, 2/3\pi)$ and quickly becomes flipped out of it.

In Fig.~\ref{BB1_fig} we illustrate the resulting probability after applying the QSP sequence with the trivial  phases  for the Chebyshev polynomials in panel (a) and BB1 phase sequence in panel (b). A comparison is made between these results  (shown with solid curves)  and those obtained from running the corresponding QSP sequences on  an  IBM-Q cloud quantum computing on the QPU based on the IBM Eagle r3 architecture \cite{AbuGhanem2025}, shown by color dots.  We obtained good agreement for this single-qubit technique despite  the  dissipative and noisy environment of real qubits in the IBM-Q processor.


The unitary operator $U_{\boldsymbol{\Phi}, a}$ can then be used in quantum simulations \cite{HamiltonianSimulation, martyn2021grand} to address computational or physical problems. In general, by varying the number of phases and their configurations using the QSP method \cite{martyn2021grand}, it is possible to construct any symmetric signal profile sensitive to the desired combinations of rotation angles.

Later, in Sec.~\ref{QSP_AIM_mapping}, we demonstrate how closely this technique resembles the adiabatic-impulse model.

\subsection{Adiabatic-impulse model}
\label{Adiabatic_impulse_Sec}
The adiabatic-impulse model  \cite{Damski2006,Higuchi2018,Heide2021,kofman2023majorana,Kofman2024,Reparaz2025} describes the evolution of a system driven by a periodic signal with a linear transition through an anti-crossing point (point of the minimal energy gap). This model works well with systems with a small energy gap and can be used as a basis for quantum logic gates \cite{Campbell2020,Caceres2023,Ryzhov_2024,Kivelae2024,Hsiao2025}. The evolution is divided into two regimes: the adiabatic regime, where the system closely follows its adiabatic states, and the diabatic impulse-type regime, where a transition of occupation probability between two energy levels occurs, accompanied by a phase gain between different energy level occupations. 

First, we consider a typical qubit Hamiltonian
\begin{align} \label{ham}
	H = \dfrac{\Delta}{2}X + \dfrac{\varepsilon(t)}{2}Z,
\end{align}
where $\Delta$ is the minimal energy gap between the two energy levels and $\varepsilon(t)=-A\cos\omega t$ is the driving signal with frequency $\omega$ and large amplitude $A>\Delta$. The evolution is separated into adiabatic evolution, described by the matrix $U$, and transition (impulse-type) evolution described by the matrix $N$:
\begin{subequations} 
	\begin{align} \label{accom_phase}
		&U(t_i,t_j) = 
		\begin{pmatrix}
			\exp{(- i \zeta(t_i,t_j))} & 0 \\
			0 & \exp{(i \zeta(t_i,t_j))} 
		\end{pmatrix} \\ \notag
		& = \exp{(- i \zeta Z)} = R_z(2\zeta), \\
		& N = 
		\begin{pmatrix}
			\mathcal{R} \exp{(- i \phi_{\text{S}})} & - \mathcal{T} \\
			\mathcal{T} & \mathcal{R}\exp{(i \phi_{\text{S}})}\
		\end{pmatrix} \\
		& = R_z(\phi_{\text{S}})R_x(\theta)R_z(\phi_{\text{S}}), \label{diabat_Rot} \\
		& \mathcal{T} = \sqrt{\mathcal{P}}, \hspace{1em} \mathcal{R} = \sqrt{1-\mathcal{P}}, \\
		& \phi_{\text{S}} = \dfrac{\pi}{4} + \delta(\ln{\delta} - 1) + \arg{\left[\Gamma(1 - i\delta)\right]}, \label{Stokes}   \\  \label{mod_prob}
		& \mathcal{P} = \exp{\left[-2\pi\delta\right]}, \hspace{1em} \delta = \dfrac{\Delta^2}{4\hbar v}, \hspace{1em} 
		\\&\left.v=\frac{d\varepsilon(t)}{dt}\right|_{\varepsilon = 0}= A\omega, 
\end{align}\end{subequations} 
where $\phi_{\text{S}}$ is the Stokes phase, $v$ is the speed of the anti-crossing passage, $\Gamma$ is the gamma function,  $\mathcal{T}$ is the transmission coefficient that is equivalent to a  polynomial base  $a$ in the QSP algorithmic technique. Moreover,  $\mathcal{R}$ is the reflection coefficient, $\mathcal{P}$ is the LZSM excitation probability of the qubit with a single transition from the ground state \cite{landau1965quantum,child1974stueckelberg, child2014molecular, Wilczek2014,Glasbrenner2024,Tedo2025,Glasbrenner2025}, and $\delta$ is the adiabaticity parameter. Note that the transition matrix $N$ depends on the direction of the crossing of the anti-crossing region and in the case of inverse crossing, the matrix $N^\text{inv}$ includes an additional phase $\pi$ along with the Stokes phase \cite{Ivakhnenko_2023,Kofman2024};  or in other words,  it equals the transposed transition matrix $N^\text{inv}=N^\top$. The phase, accumulated during the adiabatic evolution of the system, is:
\begin{subequations} 
\begin{align} \label{zeta_phase}
\zeta(t_i,t_j) &= \dfrac{1}{2 \hbar} \int^{t_j}_{t_i} \Delta E(t)dt, \\  
\Delta E(t) &= \sqrt{\varepsilon(t)^2 + \Delta^2}.
\end{align}\end{subequations} 


According to the AIM, the evolution of the system is adiabatic everywhere except at the LZSM transition point (or anti-crossing point), while the transition itself is considered instantaneous in the AIM, and the evolution during this time is described as diabatic. In this case, the complete evolution of the system during a single LZSM transition, from the initial state $\psi_\text{in}$ to the final state $\psi_\text{fin}$ consists of adiabatic evolution before the anti-crossing point, transition evolution at the anti-crossing point, and adiabatic evolution after the anti-crossing point:
\begin{align} \label{AIM-evolv}
	& U_{\text{LZSM}} = U(\zeta_2) N U(\zeta_1) = 
	\begin{pmatrix}
		U_{11} &  U_{12} \\
		-  U^{*}_{12} &  U^{*}_{11} \\
	\end{pmatrix},
	\\
	& U_{11} = \mathcal{R}_1 \exp{[-i(\phi_{\text{S}} + \zeta_1 + \zeta_2)]}, \notag\\
	& U_{12} = -\mathcal{T}_1 \exp{[i(\zeta_1 - \zeta_2)]}, \notag
\end{align}
where $\zeta_1$ and $\zeta_2$ are the phase gains before and after the anti-crossing point, respectively.
The complete evolution matrix for a single transition in terms of Bloch sphere rotation operators is
\begin{align} 
	& U_{\text{LZSM}_{}}  = R_z(2\zeta_2 + \phi_{\text{S}})R_x(\theta)R_z(2\zeta_1 + \phi_{\text{S}}).\label{LZSM_Rotations}
\end{align}
The principal structure of the AIM evolution in Eq.~\eqref{LZSM_Rotations} is identical to that of the QSP in Eq.~(\ref{QSP_rot}), differing only in parameter values that can be tuned as needed.



\section{An algorithm for converting QSP parameters to AIM}
\label{QSP_AIM_mapping}
Here, we present a method for aligning the phases of the QSP sequence with those in AIM. We develop an algorithm to adapt QSP phases for AIM, which provides the experimental inputs—amplitude, frequency, and timings—needed for the direct realization of the QSP sequence on an experimental device with LZSM-type evolution \cite{Campbell2020}, using the method of determining  driving  parameters  for  AIM \cite{Ryzhov_2024}.

  The harmonic signal $\varepsilon(t)=-A\cos\omega t$ is  linear in the vicinity of the anti-crossing region, meaning it corresponds to analytical solutions for the probability of an LZSM transition and results in the LZSM transition probability. From Eq.~(\ref{mod_prob}) we have
\begin{align}
	\left. \frac{d\varepsilon}{dt}\right|_{\varepsilon = 0} = A \omega = \dfrac{-\pi\Delta^2}{2\hbar\ln{\mathcal{P}}}.
\end{align}
The total phase gain during the LZSM transition is called the Stückelberg phase and it is expressed as the sum of the Stokes phase $\phi_{\text{S}}$ given by Eq.~(\ref{Stokes}), gained during diabatic evolution, and the phase accumulated during the adiabatic evolution of the system $\zeta(t_i,t_j)$ given by Eq.~(\ref{zeta_phase})
\begin{align}
	\phi_{\text{St}} = \phi_{\text{S}} + \zeta(t_i,t_j) . \hspace{1em} 
\end{align}

The steps for determining the amplitude $A$, frequency $\omega$, and time $\Delta t = t_j - t_i$ for a driven signal with an LZSM transition can be described as follows:

1. Given that the system undergoes a rotation by an angle $\theta$ about the $x$-axis, the probability of finding the system in the upper state after a single LZSM transition can be found from the following expression:
\begin{align} \label{prob_sin}
	\mathcal{P} = \sin^2(\theta/2).
\end{align}
Next, since we start and finish the transition far from the anti-crossing point, we fix the amplitude to ensure that the next transition starts from the same detuning $\varepsilon$ as when we finish the previous one. Using the previously defined LZSM probability in Eq.~(\ref{mod_prob}), we obtain the driving frequency:
\begin{align}
	\omega = \dfrac{-\pi\Delta^2}{2\hbar A\ln{\mathcal{P}}}.
	\label{Driving_Frequency}
\end{align}

2. The phase correspondence condition can be found by comparing the evolution matrices for a two-level system described by AIM in Eqs.~(\ref{accom_phase}, \ref{diabat_Rot}, \ref{AIM-evolv}) and the one for QSP with corresponding phases in Eq.~(\ref{QSP_evolv}):
\begin{align}
	U_{\boldsymbol{\Phi}, a} &= \exp{(i\phi_0 Z)} \prod_{k=1}^{d} W(a) \exp{(i\phi_k Z)} \\ 
	& = R_z(-2\phi_0) \prod_{k=1}^{d} R_x(-\theta) R_z(-2\phi_k). \notag
\end{align}

\begin{figure}[t] 
	\centering
	\includegraphics[width=0.5\textwidth]{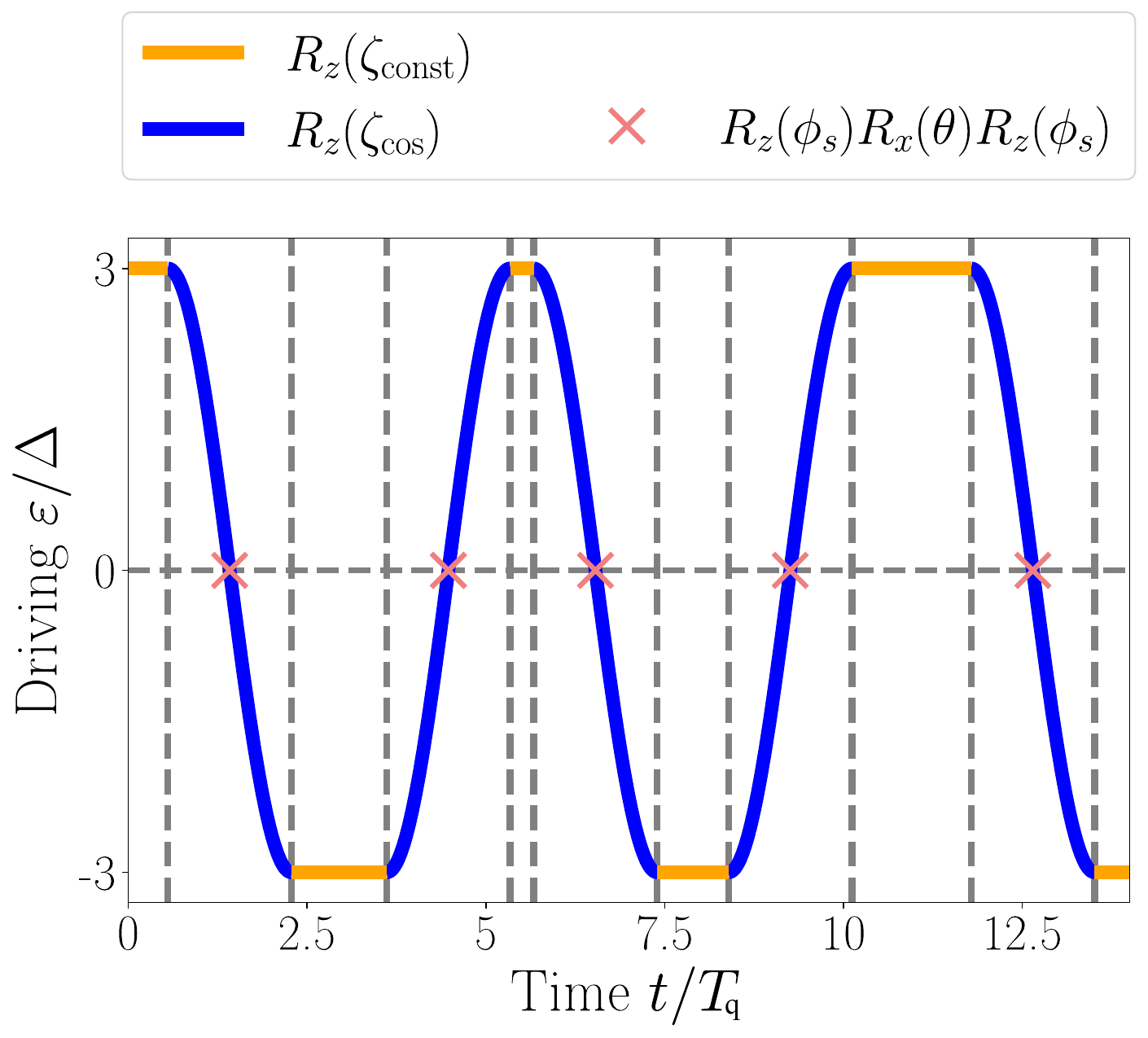}
	\caption{Signal shape for the BB1 sequence realized with AIM  and a  large amplitude  $A>\Delta$. The blue line corresponds to the   harmonic-driven  evolution described by the cosine in Eq.~(\ref{epsilon_sequence}), the orange line corresponds to the adjustable phase gain while the driving is constant, which is used to match the QSP phases with the AIM phases, and the red crosses are the transition points. 
		Time is normalized to the qubit resonant excitation period $T_\text{q}={2\pi}/{\Delta}$.}\label{signal_fig}
\end{figure}

3. We use a signal, shown in Fig.~\ref{signal_fig} and  defined as 
\begin{align} \label{epsilon_sequence}
	\varepsilon(t) = 
	\begin{cases} 
		A, & \text{if } t < T_\text{s}, \\
		-A\cos\big(\omega (t - T_\text{s})\big), & \text{if } T_\text{s} \leq t < T_\text{s} + \frac{\pi}{\omega}, \\
		- A, & \text{if } t \geq T_\text{s} + \frac{\pi}{\omega},
	\end{cases}
\end{align}
where $T_\text{s}$ is the time needed to accumulate the correct phase gain before the transition, because we use a symmetrical driving signal.

According to Step 2, for \(d\) LZSM transitions, the mapping of AIM phases to QSP is determined by the following phase relations:

\begin{align} \label{cond0}
	& \phi_{\text{S}} + 2\zeta_{\text{const}, 0}(t)+ 2\zeta_\text{cos}(t) + 2\pi j = -2\phi_0,  \notag  \\ 
	& 2 \phi_{\text{S}} + 2\zeta_{\text{const}, 1}(t) + 2\zeta_\text{cos}(t) + \pi +  2\pi j = -2\phi_1,  \notag \\ 
	& ...  \notag \\  
	& 2 \phi_{\text{S}} + 2\zeta_{\text{const}, k}(t) + 2\zeta_\text{cos}(t) + \pi +  2\pi j = -2\phi_{k},  \\
	& ... \notag \\  
	& \phi_{\text{S}} + 2\zeta_{\text{const}, d}(t) + 2\zeta_\text{cos}(t) + \pi  +  2\pi j = -2\phi_d, \hspace{0.2em} d\text{ is even,}   \notag  \\
	& \phi_{\text{S}} + 2\zeta_{\text{const,}d}(t) + 2\zeta_\text{cos}(t) + 2\pi j = -2\phi_d, \hspace{1em} d\text{ is odd.}  \notag   \label{condn} 
\end{align}

Here $j$ is an integer and $k=0,1,...,d$,
\begin{subequations} 
	\begin{align}
		&\zeta_{\text{const},k} = \Omega_\text{L} T_{\text{const},k}/2, \\ &\Omega_\text{L} = \sqrt{A^2 +\Delta^2}/\hbar, \\
		&\zeta_\text{cos} = \dfrac{1}{2\hbar}\int_{0}^{T_\text{cos}/2}  \sqrt{A^2 \cos^2\big(\omega (t - T_\text{s})\big)+\Delta^2}dt,
\end{align}\end{subequations} 
where $T_{\text{const},k}$ is the duration of the free evolution stage,  $T_\text{cos}=2\pi/\omega$ is the period of the resonance driving. 

Note that free evolution precession around the $z$-axis  happens with the Larmor frequency $\Omega_\text{L}$ and period $T_\text{L}=2\pi/\Omega_\text{L}$.
In these conditions, the phases \(\zeta_\text{const}\) and \(\zeta_\text{cos}\) are gained during the free-evolution stage, where $\varepsilon=\pm A$, and during the LZSM transition, where $\varepsilon=\pm A\text{cos}(\omega t)$, respectively. After determining the target phases \(\boldsymbol{\Phi}~=~(\phi_0, \phi_1, \dots, \phi_d)\), the evolution time is obtained to satisfy the conditions in Eq.~(\ref{cond0}) and achieve the target QSP parameters mentioned above.

 To summarize, in  this section, we  demonstrated  the direct  analogy  of the QSP technique evolution using AIM. The execution time of the operations is determined by the LZSM probability in Eq.~(\ref{mod_prob}), which corresponds to the rotation angle around the $x$-axis on the Bloch sphere and the chosen amplitude of the driving signal $A$. Notably, for specific values of the LZSM probability $\mathcal{P} = 0$, the operation time $T_\text{cos}=2\pi/\omega$, see Eq.~\eqref{Driving_Frequency} diverges to infinity. The next section is devoted to a way to resolve this problem.

 \section{quantum signal processing \\ based on double Landau-Zener-Stückelberg-Majorana transitions} 
\label{Double_LZSM}
Here we describe how to use double LZSM transitions for realizing QSP.  This allows  eliminating the huge time variability of each QSP rotation around the $x$-axis. \textit{A double LZSM transition is fully equivalent to a Mach-Zehnder interferometer with two beam splitters} (LZSM transitions) with a phase gain between beam splitters or LZSM transitions and interferometry between states (adiabatic evolution) \cite{Oliver2005,Ivakhnenko_2023,Kuzmanovic2024}. Here we use $\mathcal{P}=0.5$, which is equivalent to a (commonly used in photonics)  50/50 beam splitter. 

In particular, the Mach-Zehnder interferometer in photonic quantum computers is used as a \textit{controllable quantum gate}, controlled by adjusting the phase gain between beam splitters \cite{Bogaerts2020,Dong2023,Heide2024}.
To adjust the rotation angle of the signal operator $W$, we use an adjustable additional free evolution between two LZSM transitions.
For a double LZSM transition \cite{Ivakhnenko_2023} with adjustable phase gain between two transitions, we have 
\begin{eqnarray}
	\Xi&=& U_2 N_2 U_2 U_1 N_1^\text{inv} U_1 \notag \\ &=& U_\text{LZSM(2)} U^\text{inv}_\text{LZSM(1)}=
	\begin{pmatrix} 
		\Xi_{11}& \Xi_{12}\\
		-\Xi_{12}^* & 
		\Xi_{11}^* \end{pmatrix},
	\label{TwoPassagesMatrix}
\end{eqnarray}
where for the adiabatic evolution between transitions we use $U_2$ with an adjustable phase,  $U_2=U(\zeta_2+\phi_\text{ad})$,  where  $\phi_\text{ad}$ is an additional phase gain between two LZSM transitions. For the Stückelberg phase $\phi_\text{St}=\phi_\text{S}+\zeta_2$, we choose conditions for the X gate from Ref.~\cite{Ryzhov_2024}, $\phi_\text{St}=\frac{\pi}{2}$, $\mathcal{R}=\mathcal{T}={\frac{1}{\sqrt{2}}}$.
Then we obtain
\begin{subequations}
\begin{align}
	&\Xi_{11}=[\mathcal{R}^2 \exp{(-i (2\phi_\text{S}+2\zeta_{2}+2\phi_\text{ad}))} + \mathcal{T}^2 ] \\
	&\cdot \exp{[i (\phi_\text{ad}+\zeta_{2}-\phi_\text{in}-\zeta_{1}-\phi_\text{fin}-\zeta_{3})]}, \notag \\
	&\Xi_{12}=\mathcal{R}\mathcal{T}[1  - \exp{(-i (2\phi_{\text{S}}+2\zeta_{2}+2\phi_\text{ad}))}] \\
	&\cdot \exp{[i (\phi_{\text{S}}+\phi_\text{ad}+\zeta_{2}+\phi_\text{in}+\zeta_{1}-\phi_\text{fin}-\zeta_{3})]},\notag	
\end{align}
\end{subequations}
where
\begin{align}
	\notag	&\zeta_{1} = \zeta(0,t_{S1}),\\
	&\zeta_{2}+\phi_\text{ad} = \zeta(t_{S1},t_{S2}),\\
	&\zeta_{3} = \zeta(t_{S2},t_{\text{final}}). \notag
\end{align}	
We use a symmetric harmonic signal to perform LZSM transitions, such that $\zeta_2=\zeta_{1}+\zeta_{3}$ and $\zeta_{1}=\zeta_{3}$. Additionally, we have a phase gain before the transition $\phi_\text{in}$ and after the transition $\phi_\text{fin}$, where $t_{S{1,2}}$ represent the times at which the first and second LZSM transition signals start within the double LZSM transition (each transition signal is half period of the harmonic signal), and $t_\text{final}$ is the final time of the second LZSM transition.

\begin{figure}[t] 
	\centering
	\includegraphics[width=0.5\textwidth]{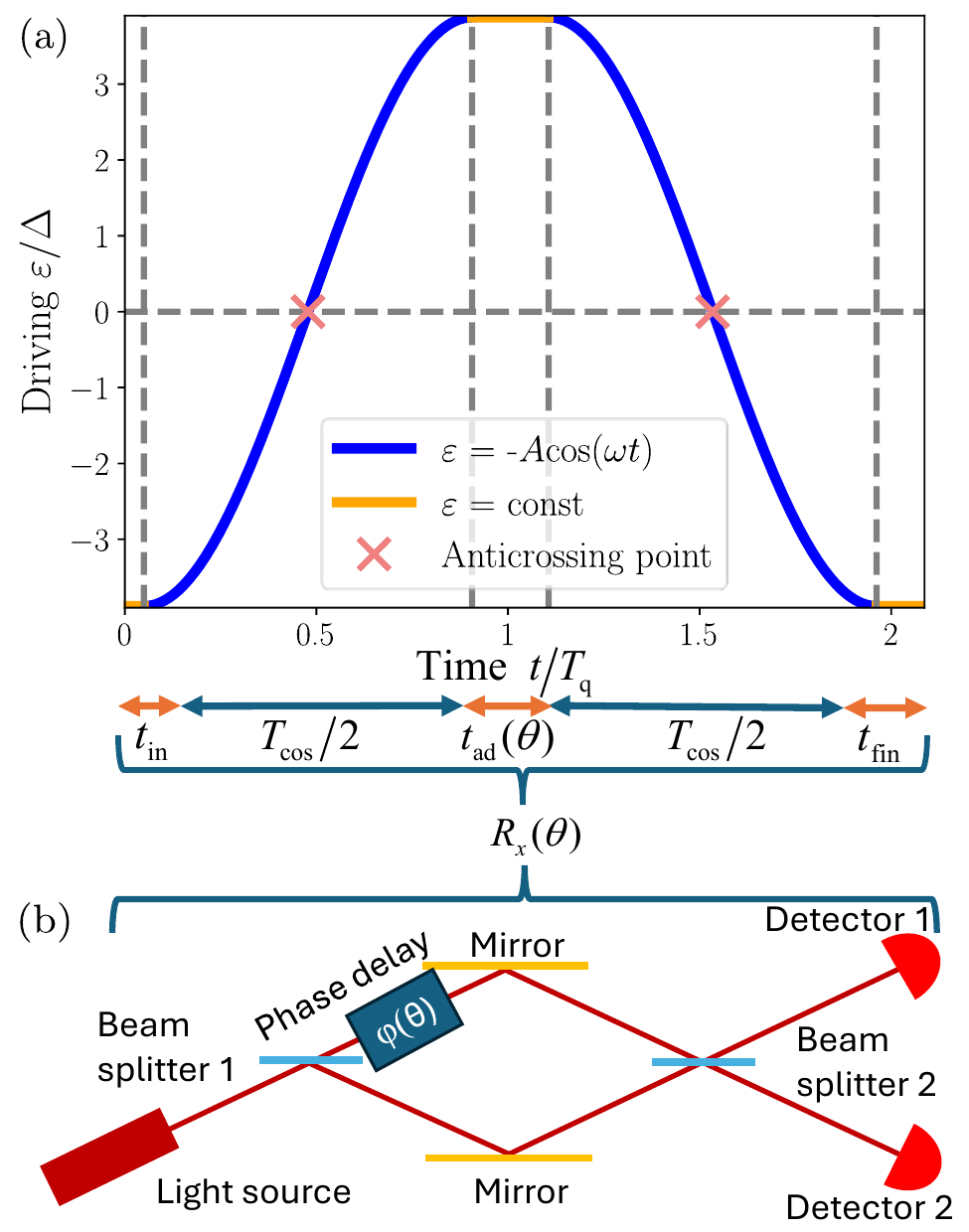}
	\caption{ (a) Adjustable rotation gate $R_x(\theta)$ with double LZSM transition. (b) Analogy for realizing an $R_x(\theta)$ rotation gate with a Mach-Zehnder interferometer.  }\label{Fig_3_Double_trans_Scheme}
\end{figure}

Next, to obtain the parameters for rotation around the \textit{x}-axis on a Bloch sphere, we compare the double LZSM transition matrix elements with the respective ones for a rotation operator $R_x(\theta)$ on a Bloch sphere
\begin{subequations} 
	\begin{align}
		&R_x(\theta)_{11}=\cos\theta/2, \\
		& R_x(\theta)_{12}=-i\sin\theta/2.
\end{align}\end{subequations} 
Then to obtain the phases, needed to perform the $R_x(\theta)$ gate, we use the same driving parameters as in  the  X gate in Ref.~\cite{Ryzhov_2024} and obtain the corresponding system of equations $\Xi=R_x(\theta)$:
\begin{align}
	\begin{cases}
		\cos\frac{\theta}{2}=\frac{1}{2}\left(1+e^{-i(\pi+2\phi_\text{ad})}\right)e^{i(\phi_\text{ad}-\phi_\text{in}-\phi_\text{fin})}, \\[2ex]	
		e^{i3\pi/2}\sin \frac{\theta}{2}=\frac{1}{2}\left(1-e^{-i(\pi+2\phi_{\text{ad}})}\right)e^{i(\phi_{\text{ad}}+\phi_{\text{in}}-\phi_{\text{fin}}+\pi/2)}.
	\end{cases}
\end{align}
As a result, by using Euler's formula, we obtain a system of equations for the phases.
By solving this system of equations, we obtain additional phases to perform the $R_x(\theta)$ operation
\begin{equation}
	\begin{cases}
		\phi_{\text{in}}=\pi/4+2\pi n,\\
		\phi_{\text{fin}}=5\pi/4+2\pi n,\\
		\phi_{\text{ad}}=\theta/2+{\pi/2}+\pi n.
	\end{cases}
\end{equation}
The minimum time of free evolution is equal to the fractional part of the phase divided by $2\pi$ and multiplied by the period of the free evolution. Thus, for all additional phases, the waiting time becomes
\begin{equation}
	\begin{cases}
		t_\text{ad}=\left\{\dfrac{\theta+\pi}{2\pi}\right\}T_\text{L}, \\
		t_\text{in}=\left\{\dfrac{\pi/2}{2\pi}\right\}T_\text{L}=\dfrac{T_\text{L}}{4},\\
		t_\text{fin}=\left\{\dfrac{5\pi/2}{2\pi}\right\}T_\text{L}=\dfrac{T_\text{L}}{4}.
	\end{cases}
\end{equation}
With the driving signal $\varepsilon(t)=-A\cos \omega t$, the first transition will be reversed,  and leads  to an additional $\pi$ phase to the initial phase gain \cite{Ivakhnenko_2023}. This gives the free evolution time that should be added at the beginning of the sequence, denoted as $t_\text{in}$, and between the signal operators, denoted as $t_\text{mid}$ for the rotation around the $x$-axis
\begin{eqnarray}
	t_\text{mid}=\left\{\frac{5\pi/2+\pi/ 2}{2\pi}\right\}T_\text{L}=\frac{T_\text{L}}{2}. 
\end{eqnarray}

\begin{figure}[h!] 
\centering
\includegraphics[width=0.5\textwidth]{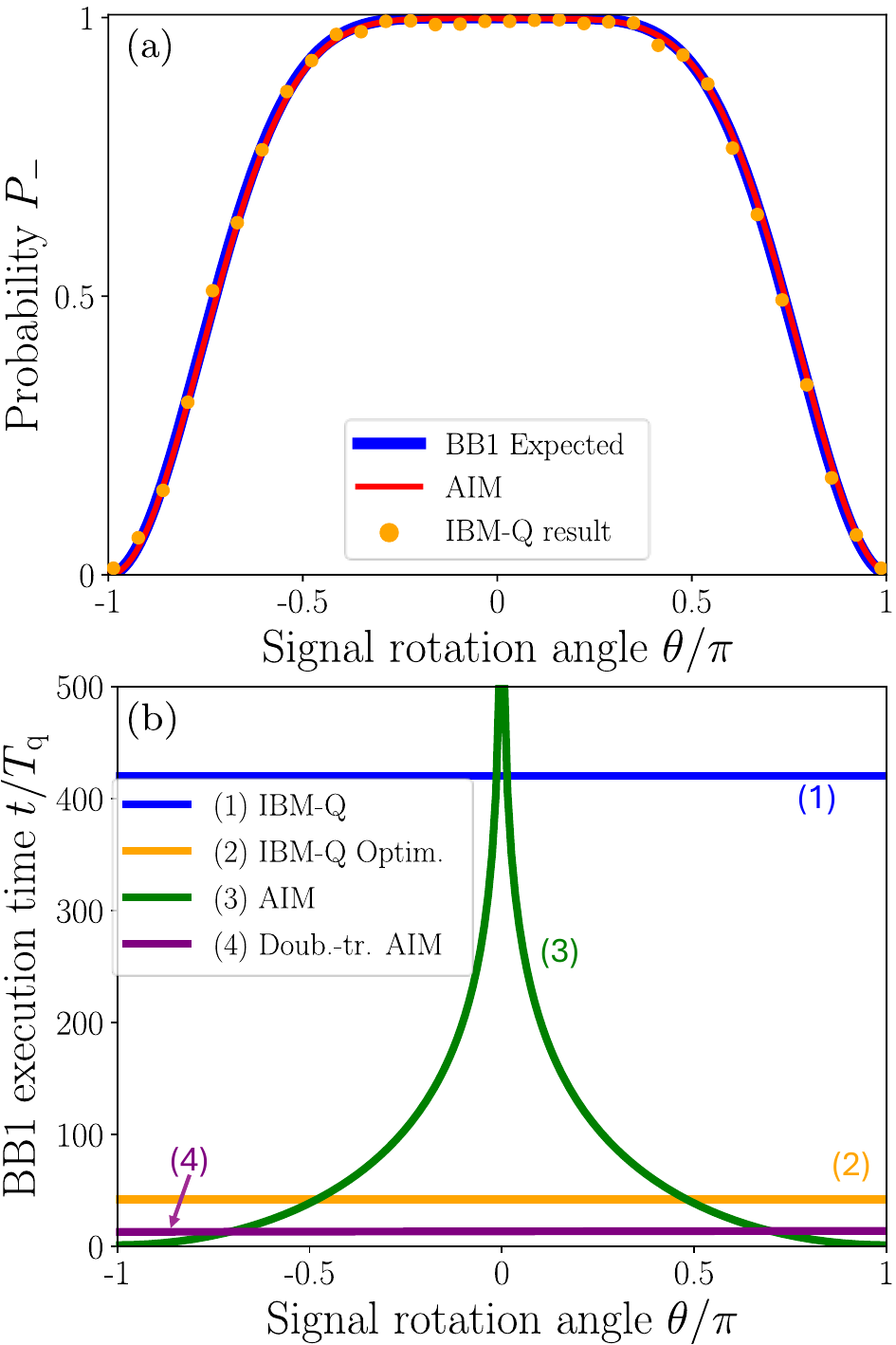}
\caption{Panel~(a) shows the resulting occupation probability $P_-$ for the BB1 sequence in Eq.~\eqref{BB1Plynom} compared to the qubit dynamics simulation result with the driving signal defined by AIM. Panel~(b) presents a comparison of the hardware execution time  $t/T_\text{q}$  for the BB1 sequence using different approaches:  (1) IBM-Q time (blue line), computed as the difference between the execution time with zero gates and the time with the BB1 sequence. This isolates the time required solely for the QSP technique, enabling direct comparison with other methods. (2) IBM-Q with  the optimization  (orange line), where the execution time does not depend on the number of single-qubit gates—i.e., the entire BB1 sequence is replaced with a single compiled gate. (3) Direct realization of QSP via AIM (green curve). (4) Double LZSM transitions (purple line). To compare  execution  times we divide all of them by the  corresponding  period of the  qubit  resonance driving $T_\text{q}={2\pi}/{\Delta}$.} \label{Fig_4_Double_trans_Result}
\end{figure}
So, we described how to get all the necessary phases, such as the phase gain time-intervals before the sequence, between double LZSM transitions, and after the sequence. To  obtain  the amplitude and the frequency, we use the algorithm described in Ref.~\cite{Ryzhov_2024} for the X gate. With these, we have all the parameters required to implement QSP sequences using double LZSM transitions. 

Double LZSM transitions allow us to eliminate the main weakness of the direct implementation of the QSP technique with AIM: the transition time diverges as the signal rotation angle approaches zero, making the result highly susceptible to decoherence and dissipation. Double transitions allow to perform the QSP sequence fast and with high fidelity by leveraging the LZSM gates technique, using the same driving signal defined for the X gate with the additional free evolution time between transitions $t_\text{ad}$ for the signal operator $W=R_x(\theta)$.
Then we  apply  the signal processing operator $S=R_z(\phi_i)$
after each transition. By choosing a working point with a higher amplitude, we can significantly improve the fidelity of LZSM gates by slightly decreasing the gate execution time.

Figure~\ref{Fig_3_Double_trans_Scheme}(a) shows the driving signal used to implement the $R_x(\theta)$ gate, where the time between transitions $t_\text{ad}$ is defined by the rotation angle around the $x$-axis, and the time intervals $t_\text{in}$ and $t_\text{fin}$ are used to gain additional phases before and after transitions to perfectly match our double LZSM transition matrix to an $R_x(\theta)$ gate. Figure~\ref{Fig_3_Double_trans_Scheme}(b) shows the equivalent Mach-Zehnder interferometer scheme, which is used as a controllable quantum logic gate for photons, where the rotation angle around the $x$-axis is defined by additional phase gain in one of the optical ways of the Mach-Zehnder interferometer, which  can be tunable.
\section{Execution time of the QSP algorithm}
Figure~\ref{Fig_4_Double_trans_Result}(a) shows a comparison of the excitation probability after applying the BB1 sequence. The IBM-Q result is shown by orange dots. Qubit dynamics simulations  were  performed with QuTiP \cite{qutip,Johansson2013,Lambert2024}, including relaxation and dephasing  with similar rates to IBM-Q, for signals defined by AIM using both direct QSP analogy and double LZSM transitions are shown by the red curve. All results show good agreement with the expected polynomial distribution shown by the blue curve. 

In Fig.~\ref{Fig_4_Double_trans_Result}(b), we compare the time needed to perform the BB1 sequence by different methods. Note that for the angle $\theta=0$ in the QSP technique, with direct QSP to AIM analogy, time diverges, as discussed before. This divergence arises due to the definition of the driving frequency in Eq.~\eqref{Driving_Frequency} and the probability of the LZSM transition Eq.~\eqref{mod_prob}, which yields $\omega=0$, so the period of the driving tends to infinity in this case. Note also that the BB1 execution time by the double LZSM transitions results in a stable operation time for all angles~$\theta$. 

We found that the IBM-Q cloud-based quantum computer needed significantly more time to execute the BB1 sequence, especially for the non-optimized version, where the dependence on the number of gates appears. This is mainly due to the Rabi-like resonance pulse type of the driving used in current quantum  processors, which loses fidelity significantly with faster gates. Despite the much faster gate execution time of the optimized algorithm  with transpilation  on an IBM-Q (the sequence of all single-qubit gates is replaced by one gate), the double LZSM transitions  are  still several times faster. But the LZSM transitions require a small energy gap in the anti-crossing point, and the distance between levels should grow fast with detuning $\varepsilon$. Therefore, the speed advantage of the LZSM-based gates may be constrained by hardware limitations in practical implementations. 

 
\section{Scalability to High-Dimensional Hilbert Spaces}
\begin{figure}[t] 
  \centering
  \includegraphics[width=0.5\textwidth]{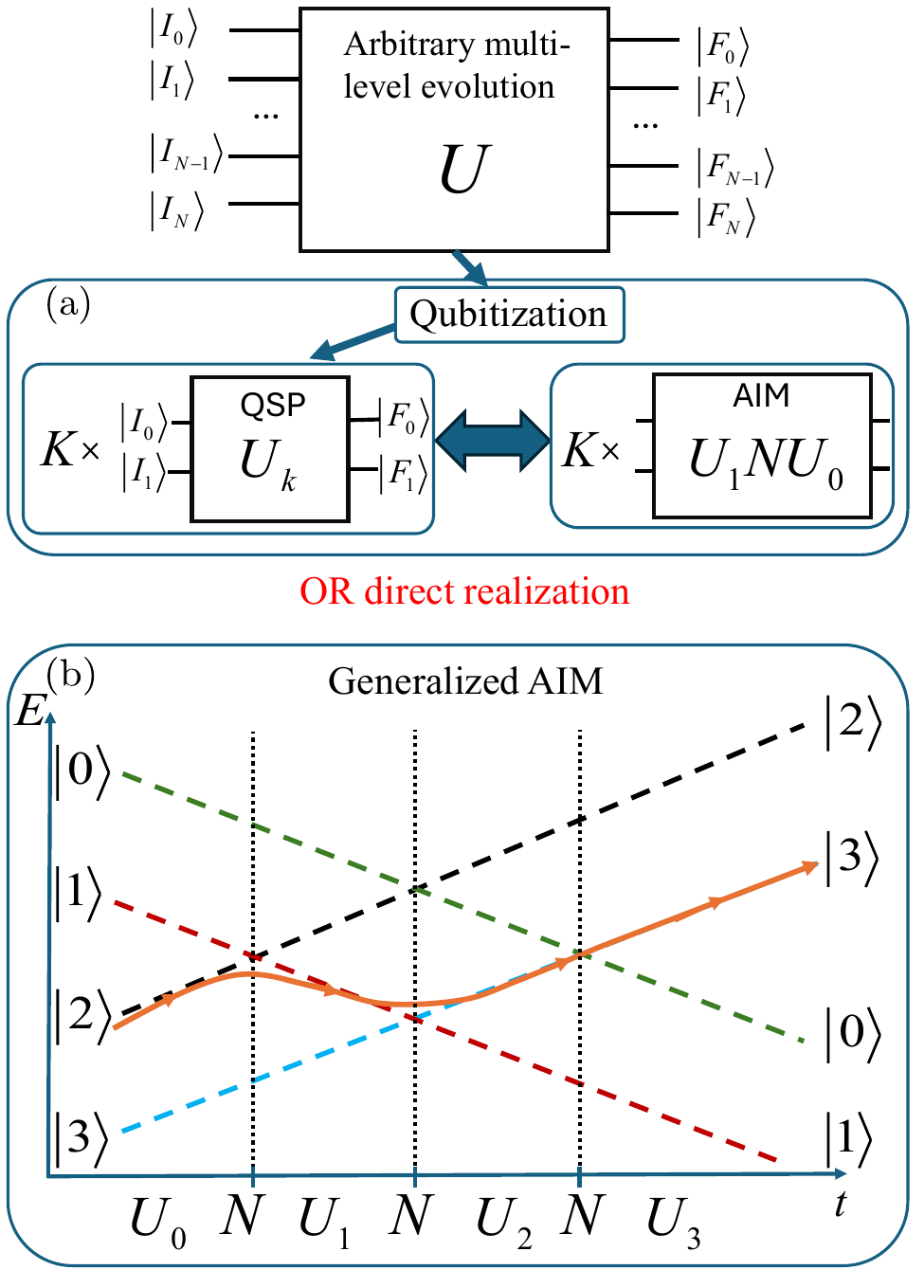}
  \caption{ Schematic diagrams for two ways to apply the adiabatic-impulse model for multi-level systems. Panel (a) shows the qubitization technique, which allows decomposing a multi-level system by $K=N(N-1)/2$ single-qubit evolution algorithms. The decomposed evolution operator can be realized with the QSP sequences, and then fast LZSM gates can be applied to perform QSP. Panel (b) demonstrates a way to directly realize multi-level evolution with generalized AIM, which consists of consecutive LZSM transitions with respective energy levels. Dashed lines represent diabatic energy levels in the two-qubit system; the orange line represents one of the possible trajectories for the occupation probability with transitions between levels on each anti-crossing point.}\label{Fig_Multilevel}
\end{figure}

To perform real-world quantum calculations dynamics of multi-level systems is crucial. While single-qubit dynamics is relatively simple, adding more levels increases the difficulty of the problem as $2^N$ for $N$ levels. In this section, we present two ways to expand our theory to multi-level systems. The first one consists in using the qubitization technique to decompose a multi-level system into many two-level systems with QSP-like evolution in each one. The second one is to directly apply the generalized AIM to a multi-level system. Both procedures are schematically presented in Fig.~\ref{Fig_Multilevel}.

\subsection{From quantum signal processing to quantum singular value transformation}
The algorithm we discussed in Sec.~\ref{QSP_AIM_mapping} can also be used for large qubit registers or multi-level systems via decomposing the multi-level problem to a certain number of single-qubit problems with the procedure known as the \textquotedblleft qubitization" technique \cite{martyn2021grand,Rossi2025}.
While QSP is often introduced in the setting of a single-qubit subspace, the same idea extends naturally to higher-dimensional systems. For a block-encoded operator acting on many qubits, the qubitization decomposes the global Hilbert space into a direct sum of two-dimensional invariant subspaces, each labeled by a singular value. For details see Appendix~\ref{QSVT_appendix}. To describe such a system, one has to \\
\begin{itemize}
\item construct a block-encoding of the operator;  
\item apply the qubitization to identify the collection of effective qubit subspaces; 
\item implement the QSP phase sequence simultaneously across all of them. 
\end{itemize}
In this way, QSP and its generalization QSVT perform polynomial transformations on the full operator spectrum, with complexity determined by the polynomial degree rather than the physical dimension of the system. This makes the framework equally applicable to large qubit registers or multi-level (qudit) systems. And can also be used to simulate an arbitrary quantum Hamiltonian with QSP sequences on quantum processors, since these usually have a lot of constraints in the parameters of the driving signal that does not allow to apply an arbitrary Hamiltonian directly.

\subsection{Generalized AIM}
Quantum logic gates based on LZSM transitions, as described in Sec.~\ref{Adiabatic_impulse_Sec} for single-qubit operations, can also be applied to multi-qubit operations \cite{Suzuki2022} with several approximations and constraints. For this, one has to
\begin{itemize}
\item locally consider each anti-crossing on the energy levels time evolution as a two-level system; 
\item apply the transition matrix that corresponds to the two respective levels for each anti-crossing; 
\item describe other levels' dynamics as the adiabatic evolution.
\end{itemize}
This is schematically described in Fig.~\ref{Fig_Multilevel}(b).
The main limitation of this approach consists in that the distance between levels should increase enough to finalize the transition process before the next transition starts. If several anti-crossings happen simultaneously between different pairs of levels, they are treated as two transitions between respective levels. The main challenge of this approach is to ensure that the relative phase between all the levels corresponds to a desired multi-level quantum gate.

\section{Conclusions}
Quantum signal processing is a powerful tool for implementing polynomial transformations and other quantum algorithms with high precision. Due to its structural similarity to the adiabatic-impulse model, it is possible to establish a direct mapping between the parameters of these two approaches, enabling efficient translation of one algorithmic technique into the other. This analogy provides a novel perspective for the direct implementation of QSP in quantum systems. 

Furthermore, the use of  the  adiabatic-impulse model, with double Landau-Zener-Stückelberg-Majorana transitions, demonstrates significantly reduced time for implementing  the  quantum signal processing technique. The double LZSM transitions allow us to solve the infinite transition time problem for $\mathcal{P}=0$ in the direct implementation of QSP with AIM. 

The double LZSM transitions approach is a full analog of the Mach-Zehnder interferometer, which allows to use the double transition approach from AIM in a controllable quantum gate for quantum photonic computations, so the same approaches can be used for both of them. With the double LZSM transition, we show how to perform rotations around the $x$-axis on the Bloch sphere for any desired angle $\theta$. By combining this with idling times to perform signal processing operators, we can use fast non-adiabatic LZSM gates to perform any QSP sequence and  achieve  faster execution time even for the most optimized case for the IBM-Q, where all single-qubit gates are combined into one. In principle, the same single-qubit gate sequence combination technique can be applied for LZSM gates to reduce the execution time accordingly.

The adiabatic-impulse model describes well the dynamics of systems with small energy gaps and can be used as an alternative approach to implement fast quantum logic gates and algorithmic techniques such as QSP.

\section*{ACKNOWLEDGMENTS}
The authors acknowledge the useful discussions with Z. M. Rossi. S.N.S. acknowledges support via the
Virtual Short-Term Scientist Engagement Fellowship Program for Ukrainian Scientists with Dual-Use Relevant Expertise (STCU, Grant No. NSEP1030). F.N. is supported in part by the Japan Science and Technology Agency (JST) [via the CREST Quantum Frontiers program Grant No. JPMJCR24I2, the Quantum Leap Flagship Program (Q-LEAP), and the Moonshot R\&D Grant No. JPMJMS2061], and the Office of Naval Research (ONR) Global (via Grant No. N62909-23-1-2074).
\section*{DATA AVAILABILITY}
The data that support the findings of this article are not publicly available. The data are available from the authors upon reasonable request. \nocite{apsrev41Control} 

\appendix

\section{Interpretation of the quantum singular value transformation in terms of rotations}
\label{QSVT_appendix}

The framework of quantum signal processing (QSP) admits a natural generalization to the quantum singular value transformation (QSVT)~\cite{martyn2021grand}. In this setting, arbitrary matrices can be embedded into larger unitaries via block-encoding, which then allows one to implement polynomial transformations of their singular values. This embedding procedure is called \emph{qubitization}, since it reduces the dynamics of a potentially large Hilbert space to effective two-level systems.

Every complex matrix $C$ admits a singular value decomposition:
\begin{equation}
    C = \sum_{i=1}^r \sigma_i \ket{w_i}\bra{v_i},
\end{equation}
where $r = \mathrm{rank}(C)$, $\sigma_i \geq 0$ are the singular values of $C$, $\{\ket{v_i}\}$ are the right singular vectors, which span the input space of $C$, and $\{\ket{w_i}\}$ are the left singular vectors, which span the output space of $C$. QSVT is designed to act directly on these singular values by implementing polynomial transformations $\sigma_i \mapsto P(\sigma_i)$.

The appearance of singular values is natural in block-encoding because the input and output subspaces of the encoded operator may differ in dimension. Unlike unitary operators (where eigenvalues suffice), a general matrix must be described in terms of its singular value decomposition. \\

Suppose we wish to encode $C$ into a larger unitary $U.$ Then there exist projectors $\Pi$ and $\widetilde\Pi$ such that
\begin{equation}
    C = \widetilde\Pi U \Pi.
\end{equation}
The meaning of these projectors is that $\Pi$ selects the right singular vector space (the input block of $U$) and $\widetilde\Pi$ selects the left singular vector space (the output block of $U$). Thus, $\Pi$ identifies where inputs are injected, and $\widetilde\Pi$ identifies where outputs appear. In QSVT, we also introduce phase-modified projectors:
\begin{align}
    &\Pi_\phi = \exp{[i\phi(2\Pi-I)]}, \\
    & \widetilde\Pi_\phi = \exp{[i\phi(2\widetilde\Pi-I)]}.
\end{align}
These are unitary transformations that apply $z$-axis rotations on the subspaces marked by $\Pi$ and $\widetilde\Pi$. {We adopt the convention that $\Pi$ projects onto the index state $\ket{1}$ and $\widetilde\Pi$ projects onto $\ket{0}$. With this choice, $\Pi_\phi|_{\mathcal S_i} = e^{-i\phi Z}$ and $\widetilde\Pi_\phi|_{\mathcal S_i} = e^{+i\phi Z}$. This fixes the sign conventions once and for all.} \\

The qubitization ensures that the Hilbert space decomposes into invariant two-dimensional subspaces
\begin{equation}
    \mathcal S_i = \mathrm{span}\{\ket{0}\otimes\ket{w_i},\;\ket{1}\otimes\ket{v_i}\},
\end{equation}
with one such subspace for each singular value $\sigma_i$. On each $\mathcal S_i$, the block-encoded $U$ acts as an $SU(2)$ matrix,
\begin{equation}
    U|_{\mathcal S_i} = R(\sigma_i).
\end{equation}
To make unitarity explicit, we parameterize the singular value as
\[
\sigma_i = \cos\!\left(\tfrac{\theta_i}{2}\right), \qquad \theta_i = 2\arccos\sigma_i.
\]
Then
\[
R(\sigma_i) = R_y(\theta_i) = e^{-i(\theta_i/2)Y}.
\]
Equivalently, in $X$–$Z$ language one may write
\[
R(\sigma_i) = e^{i\pi Z/4}\,R_x(\theta_i)\,e^{-i\pi Z/4}.
\]
Restricted to $\mathcal S_i$, the projector phases act diagonally:
\begin{equation}
    \Pi_\phi|_{\mathcal S_i} = \exp{(-i\phi Z)}, \qquad \widetilde\Pi_\phi|_{\mathcal S_i} = \exp{(i\phi Z)}.
\end{equation}
These represent opposite $z$-axis rotations of the effective qubit.

The central building block of QSVT is the iterate~\cite{martyn2021grand}
\begin{equation}
    \widetilde U_{\boldsymbol{\Phi}}
    =
    \prod_{k=1}^{d/2}
    \Bigl(\Pi_{\phi_{2k-1}}\,U^\dagger\,\widetilde\Pi_{\phi_{2k}}\,U\Bigr),
    \label{eq:qsvt_even}
\end{equation}
where $d$ is the degree of the implemented polynomial and $\{\phi_j\}$ are tunable phase parameters. {This form is valid when $d$ is even, in which case the input and output spaces are both the right singular vector space spanned by $\{\ket{v_i}\}$.}
{For odd $d$, the iterate has an additional prefactor and connects the right singular space to the left singular space.} 

Restricted to $\mathcal S_i$, one factor of the even-$d$ product reduces to
\begin{align}
    &\Pi_{\phi_{2k-1}} U^\dagger \widetilde\Pi_{\phi_{2k}} U \big|_{\mathcal S_i} \\
    &\qquad = \exp{(-i\phi_{2k-1}Z)}R(\sigma_i)\exp{(i\phi_{2k}Z)}R(\sigma_i). \notag
\end{align}
Here $U$ and $U^\dagger$ swap between right and left singular vector bases, while the exponential factors encode QSP phases. Then
\begin{align}
    &\sigma_i=\cos(\theta_i/2) \\
    &R(\sigma_i) = R_y(\theta_i). \notag
\end{align}
With $Z R_y(\theta) = R_y(-\theta) Z$, the full QSVT sequence becomes
\begin{equation}
    \widetilde U_{\boldsymbol{\Phi}}|_{\mathcal S_i} =
    \prod_{k=1}^{d/2} 
    \Bigl(
        R_z(2\phi_{2k-1}) R_y(\theta_i) R_z(2\phi_{2k}) R_y(-\theta_i)
    \Bigr),
    \label{eq:qsvt_rotation_form}
\end{equation}
where $\exp{(i\phi Z)}=R_z(-2\phi)$.
Finally, recalling that $R_y(\theta)$ can be rewritten using $X$-axis rotations,
\begin{equation}
    R_y(\theta) = \exp{\left(i\frac{\pi}{4}Z\right)}R_x(\theta)\exp{\left(-i\frac{\pi}{4}Z\right)},
\end{equation}
we obtain
\begin{align}
    \widetilde U_{\boldsymbol{\Phi}}|_{\mathcal S_i}
    = \exp{\left(i\frac{\pi}{4}Z\right)}
      \Biggl[
        \prod_{k=1}^{d/2}
        & \Bigl(
          R_z(2\phi_{2k-1}) R_x(\theta_i)\\ \times & R_z(2\phi_{2k}) R_x(-\theta_i)
        \Bigr)
      \Biggr]
    \exp{\left(-i\frac{\pi}{4}Z\right)}. \notag
\end{align}
 Similarly, one can get the result for the odd $d$, for which the pattern of repeatedly applied $R_xR_z$ gates is preserved. 

Thus, after qubitization, each singular subspace $\mathcal S_i$ is reduced to a qubit undergoing controlled rotations about the $Z$ and $X$ axes. The sequence of phases $\{\phi_j\}$ fully determines the polynomial transformation implemented, while the angle $\theta_i$ encodes the singular value $\sigma_i = \cos\theta_i$. 

In the special case where the block-encoded operator is itself unitary, all singular values have magnitude one, and QSVT reduces to the original QSP framework acting on eigenvalues lying on the unit circle. QSVT thus generalizes QSP by allowing polynomial transformations of singular values of arbitrary operators, not just phases of unitaries.
\bibliography{bibfile,1}
\end{document}